\documentclass[a4paper,12pt]{article}
\usepackage[cp1251]{inputenc}
\usepackage[english]{babel}

\def\bb{\begin{equation}}
\def\ee{\end{equation}}

\begin{document}

\begin{center}
{\Large Discrete Toda field equations}

\vskip 0.2cm

{Ismagil Habibullin}

{e-mail: ihabib@imat.rb.ru}
\end{center}
\vskip 0.2cm

\begin{abstract}
Discrete analogs of the finite and affine Toda field equations are
found corresponding to the Lie algebras of series $C_N$ and
$\tilde{C_N}$.
\end{abstract}

\section{Introduction}

Consider the Toda chain with three discrete independent variables
$u$, $v$, $k$ (see \cite{hir}, \cite{lps}):
\begin{equation}\label{dtoda}
e^{f_{uv}-f_u-f_v+f}=\frac{1+he^{f^1_v-f_u}}{1+he^{f_v-f_u^{-1}}}.
\end{equation}
Here $f=f(u,v,k)$ is an unknown function, and the following
notations are accepted. The upper index is used to indicate shifts
with respect to the third variable $k$ so that $f^1=f(u,v,k+1)$
and $f^{-1}=f(u,v,k-1)$. The lower index shows shifts of the first
and second variables $f_u=f(u+h,v,k)$, $f_{-u}=f(u-h,v,k)$,
$f_v=f(u,v+h,k)$, $f_{-v}=f(u,v-h,k)$, $h$ is the parameter of the
grid such that for small values of $h$ one gets $f_u-f\simeq\sqrt
hD_xf$, and $f_v-f\simeq\sqrt hD_yf$, and $f_{uv}-f_u-f_v+f\simeq
hD_xD_yf$, where $D_x$ and $D_y$ differential operator with
respect to $x$ and $y$. Evidently the continuum limit
$h\rightarrow 0$ of the discrete Toda chain (DTC) (\ref{dtoda})
gives the usual two-dimensional Toda chain \cite{dar}
\begin{equation}\label{toda}
D_xD_yf=e^{f^1-f}-e^{f-f^{-1}}.
\end{equation}
Chain (DTC) is closely connected with the well known discrete
bilinear Hirota-Miwa equation
\begin{equation}\label{hir}
tt_{uv}-t_ut_v=t_v^1t_u^{-1}.
\end{equation}
Here the unknown $t=t(u,v,k)$ depends also upon three independent
discrete variables. Miura type transformation $t^1=e^ft$ converts
the equation (\ref{hir}) into the equation (\ref{dtoda}).

The chain (\ref{dtoda}) admits evidently periodical closure
constraint $f(u,v,k)=f(u,v,k+N)$ which reduces it to a finite
field equation with dynamical variables $f(u,v,1)$, $f(u,v,2)$,
... $f(u,v,N)$. The other closure is defined by the degeneration
points of the chain. For example, the chain (\ref{dtoda})
truncated by the conditions $e^{-f(u,v,0)}=0$ and
$e^{f(u,v,N+1)}=0$ is an integrable finite system of discrete
hyperbolic type equations. But as a rule finite field reductions
of chains are not exhausted by degenerate and periodic ones. For
instance, the chain (\ref{toda}) admits a large class of
reductions connected with the simple and affine Lie algebras
\cite{ls}. Aforementioned degenerate and periodic reductions
correspond to the Lie algebras of series $A_n$ and $\tilde A_n$,
respectively. The problem of looking for discrete analogs of
finite reductions of (\ref{toda}) corresponding to the other Lie
algebras of the finite grows is still open (see, for example,
\cite{w}, and also the surveys \cite{zab} and \cite{zab1}). In
\cite{w} the examples of truncations of (\ref{dtoda}) connected
with special classes of solutions have been discussed. We suggest
below discrete analogs of the Toda chain of the series $C_N$,
$\tilde C_N$.

To shorten the formulae introduce notations
$z(u,v,k)=he^{f_v-f^{-1}_u}+1$ and $\Delta f=f_{uv}-f_u-f_v+f$,
then the chain (\ref{dtoda}) gets the form $z^{1}=e^{\Delta f}z$.
Without loosing the generality one can put $h=1$. Really, the
parameter $h$ can be made equal to one by the following shift
$f(k)\rightarrow f(k)+k\ln h$.

The main result of the paper is given in the following two
statements.

{\bf Proposition 1.} The discrete Toda field equation
\begin{eqnarray}
&&e^{\Delta f^{-1}}=\frac{z}{z^1_{-u,v}}\nonumber,\\
&&e^{\Delta f^j}=\frac{z^{j+1}}{z^j},\qquad 0\leq j\leq N-1
\label{ctoda2},\\
&&e^{\Delta f^N}=\frac{1}{z^N}\nonumber
\end{eqnarray}
admits the Lax pair (see (\ref{cpeqs})-(\ref{cqeqs}) below). In
the continuum limit $h\rightarrow 0$ the chain (\ref{ctoda2})
turns into the Toda equation of the series $C_N$:
\begin{eqnarray}
&&u(-1)=-u(0)\nonumber,\\
&&u_{xy}=e^{u(k+1)-u(k)}-e^{u(k)-u(k-1)}\label{ctoda},\\
&&e^{u(N+1)}=0\nonumber.
\end{eqnarray}

{\bf Proposition 2.} The discrete Toda field equation
\begin{eqnarray}
&&e^{\Delta f^{-1}}=\frac{z}{z_{-u,v}^1},\nonumber \\
&&e^{\Delta f^{j}}=\frac{z^{j+1}}{z^j}, \qquad
0\leq j\leq N-1,\label{tctoda2}\\
&&e^{\Delta f^{N}}=\frac{z^{N-1}_{u,-v}}{z^N},\nonumber
\end{eqnarray}
admits the Lax pair (see (\ref{fPeqs})-(\ref{fQeqs}), below). In
the continuum limit $h\rightarrow 0$ it turns into the Toda
equation of the series $\tilde{C}_N$:
\begin{eqnarray}
&&u(-1)=-u(0)\nonumber,\\
&&u_{xy}=e^{u(k+1)-u(k)}-e^{u(k)-u(k-1)}\label{tctoda},\\
&&u(N)=-u(N-1)\nonumber
\end{eqnarray}

\section{Involutions of the associated linear
systems}

The chain (\ref{dtoda}) admits the Lax pair consisting of two
linear discrete equations \cite{hir}
\begin{equation}\label{laxp}
\psi_u=e^{f_u-f}\psi-\psi^1,\qquad \psi_v=\psi
+e^{f_v-f^{-1}}\psi^{-1}.
\end{equation}
Here the indices of the eigenfunction $\psi=\psi(u,v,k)$ are for
the shifts according to the rule, defined above:
$\psi_u=\psi(u+1,v,k)$, $\psi_v=\psi(u,v+1,k)$,
$\psi^1=\psi(u,v,k+1)$, $\psi^{-1}=\psi(u,v,k-1)$ and so on.

Exclude from the system of equations (\ref{laxp}) all the shifts
of the third variable. As a result one gets a linear discrete
hyperbolic equation
\begin{equation}\label{hyp1}
\psi_{uv}-\psi_{u}-e^{f_{uv}-f_v}(\psi_{v}-z\psi)=0.
\end{equation}
Below we will need equations dual to (\ref{laxp}) and
(\ref{hyp1}). In order to find the dual equations we use the
discrete symmetries of the chain (\ref{dtoda}). Evidently the
chain is invariant under involution $u\rightarrow 1-v$,
$v\rightarrow 1-u$. However the involution changes the Lax pair
(\ref{laxp}), which takes now the form
\begin{equation}\label{laxy}
y_{-u}=y+e^{f_{-u}-f^{-1}}y^{-1},\qquad
y_{-v}=e^{f_{-v}-f}y-y^{1}.
\end{equation}
The hyperbolic equation (\ref{hyp1}) turns into the equation
\begin{equation}\label{hyp2}
y_{uv}-{1\over z}y_{v}-{e^{f_{v}-f}\over z}(y_{u}-y)=0.
\end{equation}

The idea to use two (mutually conjugate) Lax pairs when studying
the finite Toda chain belongs to the classical work by Darboux
\cite{dar}.

By analogy with the continuous case pose the question when two
hyperbolic equations (\ref{hyp1}) and (\ref{hyp2}) are related to
each other by a multiplicative transform like $y=a\psi$  (see,
also, \cite{h})?  Two hyperbolic equations are connected by a
multiplicative transform if and only if their Laplace invariants
are the same \cite{nov}. Remind that the Laplace invariants of the
equation
\begin{equation}\label{hyp}
a\psi+b\psi_{u}+c\psi_{v}+d\psi_{uv}=0
\end{equation}
are expressed as
\begin{equation}\label{laplace}
K_1=\frac{bc_u}{da_u}, \qquad K_2=\frac{b_vc}{da_v}.
\end{equation}
Denote through $K_{1\psi}(u,v,k)$, $K_{2\psi}(u,v,k)$ Laplace
invariants of the equation (\ref{hyp1}) and through
$K_{1y}(u,v,k)$, $K_{2y}(u,v,k)$ -- Laplace invariants of the
equation (\ref{hyp2}). Compute all these invariants and find
\begin{equation}\label{lap}
K_{1y}={1\over z^1},\quad K_{2y}={1\over z}, \quad
K_{1\psi}={1\over z_u},\quad K_{2\psi}={1\over z^1_v}.
\end{equation}
Therefore, if the field variables satisfy the constraint
\begin{equation}\label{cons}
z(u+1,v,k_0-1)=z(u,v+1,k_0+1)
\end{equation}
for a fixed value $k=k_0$, then the Laplace invariants of these
equations will satisfy the following conditions
\begin{equation}\label{kpsi}
K_{1y}(u+1,v,k_0-1)=K_{1\psi}(u,v,k_0),\quad
K_{2y}(u+1,v,k_0-1)=K_{2\psi}(u,v,k_0),
\end{equation}
and the conditions
\begin{equation}\label{ky}
K_{1\psi}(u,v-1,k_0-1)=K_{1y}(u,v,k_0),\quad
K_{2\psi}(u,v-1,k_0-1)=K_{2y}(u,v,k_0).
\end{equation}
Consequently, there are such functions $R=R(u,v)$, $S=S(u,v)$ that
the following relations take place
\begin{equation}\label{cons2}
\psi(u,v-1,k_0-1)=Ry(u,v,k_0) \quad\mbox{and}\quad
y(u+1,v,k_0-1)=S\psi(u,v,k_0)
\end{equation}
between solutions of the hyperbolic equations (\ref{hyp1}) and
(\ref{hyp2}). Multipliers $R$ and $S$ are found from the following
overdetermined system of linear equations
\begin{eqnarray}
\label{R} &R_u=Rz_{-v}e^{f^{-1}_{u,-v}-f^{-1}_{-v}}, \quad
&R_v=Rz_{-u}e^{f_{-u}-f_{-u,v}},\\
&S_u=S\frac{e^{f-f_u}}{z^{-1}_{-u,-v}},\qquad\qquad
&S_v=S{\displaystyle{\frac{e^{f^{-1}_v-f^{-1}}}{z^{-1}}}},
\label{S}
\end{eqnarray}
consistency of which is guaranteed by the condition (\ref{cons}).

\section{Lax pair of the semi-infinite chain}

Equations (\ref{cons2}) can be referred to as boundary conditions
cutting off the linear equations (\ref{laxp}), (\ref{laxy})
reducing them into the half-line $k\geq k_0$ (or the half-line
$k\leq k_0$). Concentrate on this statement. Put first for the
simplicity  $k_0=0$. Substituting (\ref{cons2}) into the equations
(\ref{laxp}) and (\ref{laxy}) for $k=0$ one gets
\begin{equation}\label{cons3}
y^0_{-u}=y^0+X_{-u}\psi^0_{-u}, \quad \psi^0_{v}=\psi^0+H_vy^0_v,
\end{equation}
where $X=Se^{f-f_u^{-1}}$, $H=Re^{f-f_{-v}^{-1}}$.

The following lemma gives the connection between the functions $X$
and $H$.

{\bf Lemma 1.} Solutions of the system of the equations (\ref{R}),
(\ref{S}) can be chosen to satisfy the constraint
\begin{equation}\label{HX}
HX=\frac{z_{-v}-1}{z_{-v}}.
\end{equation}

Proof. It follows from the equations (\ref{R}) and (\ref{S}) that
functions $X$ and $H$ satisfy the similar linear difference
equations of the first order
\begin{eqnarray}
\label{H} &X_v=X\frac{e^{f_v-f}}{z} \quad
&H_v=Hz_{-u}^1e^{f^{-1}_{-v}-f^{-1}},\\
&X_u=X\displaystyle{\frac{e^{f_u^{-1}-f_{u^2}^{-1}}}{z^{1}}},\quad
&X_v=X\frac{e^{f_v-f}}{z} \label{X},
\end{eqnarray}
where $f_{u^2}^{-1}:=f(u+2,v,k-1)$.

It is not difficult to check that if the condition (\ref{HX})
holds at a point $(u,v)$ then it also holds at any neighbouring
point.

Let us discuss briefly dynamical variables. Evidently shifts of
the eigenfunction $\psi$ in the positive direction: $\psi_u$,
$\psi_{u^2}$, $\psi_v$, $\psi_{v^2}$, ... as well as shifts of the
eigenfunction $y$ in the negative direction such as $y_{-u}$,
$y_{-u^2}$, $y_{-v}$, $y_{-v^2}$, ... are locally expressed trough
unshifted (dynamical) variables $\psi$, $\psi^{\pm1}$,
$\psi^{\pm2}$, ... and $y$, $y^{\pm1}$, $y^{\pm2}$, ... . This is
not the case for the shifts on the opposite directions which
really should be considered as nonlocal variables. Actually these
variables cannot be expressed through a finite number of dynamical
ones. For example, to find the variable  $\psi_{-u}$ it is
necessary to solve the difference equation (with respect to the
argument $k$) of the form
$$\psi_{-u}^1-e^{f-f_{-u}} \psi_{-u}= \psi. $$
Enlarge the set of dynamical variables. In addition to the set of
dynamical variables on the half-line $k\geq 0$, consisting of the
functions $\{\psi^0,y^0,\psi^1,y^1,... \}$, introduce two more
variables $Y$ and $\Psi$ by setting $Y=y_v^0$, $\Psi=\psi_{-u}^0$.

{\bf Lemma 2.} The shifts $Y_{-u}$, $Y_{-v}$, $Y_{u}$ of the
variable $Y$ as well as the shifts $\Psi_{u}$, $\Psi_{v}$,
$\Psi_{-v}$ of the variable $\Psi$ are linearly expressed through
a finite number of the elements of the enlarged dynamical set
$\{\Psi, Y,\psi^0,y^0,\psi^1,y^1,... \}$.

Proof. Some of the equations required follows directly from the
definition: $Y_{-v}=y^0$, $\Psi_{u}=\psi^0$. Let us shift the
first equation (\ref{cons3}) to the right by one with respect to
$u$ and to $v$ and rewrite it in the form
$$Y_u=Y-X_v\psi^0_v=Y-X_v(\psi^0+H_vY)=Y(1-X_vH_v)-X_v\psi^0.$$
Due to the Lemma 1 the quantity inside the parentheses is equal to
$\displaystyle{{1\over z}}$, hence
$$Y_u={1\over z}Y-X_v\psi^0.$$
Shift the expression obtained to the left by one respect to $u$
and transform it as
\begin{equation}\label{Y-u}
Y_{-u}=z_{-u}Y+X_{-u,v}z_{-u}\Psi.
\end{equation}
Shift now the second equation of (\ref{cons3}) by one to the left
respect $u$ and $v$ then after slight simplification one gets
$$\Psi_{-v}=\displaystyle{{1\over z_{-u,-v}}}\Psi-H_{-u}y^0,$$
and
\begin{equation}\label{Psi-v}
\Psi_{v}=z_{-u}\Psi+H_{-u,v}z_{-u}Y.
\end{equation}
Lemma 2 is proved.

The commentary to the lemma. Variables $\psi^j$ can be shifted
upward and to the right on the $(u,v)$-plane while the variables
$y^j$ -- downward and to the left. The new variables are special
-- they can be shifted on three directions.

Summarize the computations above. Introduce some notations. Denote
through $P$ and $Q$ infinite dimensional vectors-columns such as
\begin{equation}\label{PQ}
P=\left(%
\begin{array}{c}
  Y \\
  y^0 \\
  y^1 \\
 y^2  \\
  \cdots \\
\end{array}%
\right), \qquad
Q=\left(%
\begin{array}{c}
  \Psi \\
  \psi^0 \\
  \psi^1 \\
 \psi^2  \\
  \cdots \\
\end{array}%
\right),
\end{equation}
i.e. $P_0=Y$, $Q_0=\Psi$ and $Q_i=\psi^{i-1}$ for  $i\geq 1$
$P_i=y^{i-1}$. Introduce six infinite dimensional matrices, four
of which are two diagonal
\begin{equation}\label{AB}
A=\left(%
\begin{array}{ccccc}
  z_u & 0 & 0 & 0 & \cdots \\
  0 & 1 & 0 & 0 & \cdots \\
  0 & e^{f^1_{-u}-f} & 1 & 0 & \cdots \\
  0 & 0 & e^{f^2_{-u}-f^1} & 1 & \cdots \\
  \cdots & \cdots & \cdots & \cdots & \cdots \\
\end{array}%
\right), \quad
B=\left(%
\begin{array}{ccccc}
  0 & 1 & 0 & 0 & \cdots \\
  0 & e^{f_{-v}-f} & -1 & 0 & \cdots \\
  0 & 0 & e^{f^1_{-v}-f^1} & -1 & \cdots \\
  0 & 0 & 0 & e^{f^2_{-v}-f^2} & \cdots \\
  \cdots & \cdots & \cdots & \cdots & \cdots \\
\end{array}%
\right),
\end{equation}
\begin{equation}\label{CD}
C=\left(%
\begin{array}{ccccc}
  z_v & 0 & 0 & 0 & \cdots \\
  0 & 1 & 0 & 0 & \cdots \\
  0 & e^{f^1_{v}-f} & 1 & 0 & \cdots \\
  0 & 0 & e^{f^2_{v}-f^1} & 1 & \cdots \\
  \cdots & \cdots & \cdots & \cdots & \cdots \\
\end{array}%
\right), \quad
D=\left(%
\begin{array}{ccccc}
  0 & 1 & 0 & 0 & \cdots \\
  0 & e^{f_{u}-f} & -1 & 0 & \cdots \\
  0 & 0 & e^{f^1_{u}-f^1} & -1 & \cdots \\
  0 & 0 & 0 & e^{f^2_{u}-f^2} & \cdots \\
  \cdots & \cdots & \cdots & \cdots & \cdots \\
\end{array}%
\right),
\end{equation}
and two others are of the form
\begin{equation}\label{ac}
a=\left(%
\begin{array}{ccc}
  X_{-u,v}z_{-u} & 0 & \cdots \\
  X_{-u} & 0 & \cdots \\
  0 & 0 & \cdots \\
  \cdots & \cdots & \cdots \\
\end{array}%
\right),\quad
c=\left(%
\begin{array}{ccc}
  H_{-u,v}z_{v} & 0 & \cdots \\
  H_{v} & 0 & \cdots \\
  0 & 0 & \cdots \\
  \cdots & \cdots & \cdots \\
\end{array}%
\right).
\end{equation}
The last two matrices have all entries to be zero except the
following ones $a_{0,0}$, $a_{1,0}$, $c_{0,0}$, $c_{1,0}$.

Make up a system of difference equations
\begin{eqnarray}
&P_{-u}=AP+aQ,\quad &P_{-v}=BP,\label{Peqs}\\
&Q_v=CQ+cP,\quad&Q_u=DQ.\label{Qeqs}
\end{eqnarray}

{\bf Proposition 3.} The systems (\ref{Peqs}) and (\ref{Qeqs}) are
consistent if and only if their coefficients satisfy the
semi-infinite lattice of equations
\begin{eqnarray}
&&e^{\Delta f^{-1}}=\frac{z}{z_{-u,v}^1},\label{semibc}\\
&&e^{\Delta f^{j}}=\frac{z^{j+1}}{z^j}, \quad \mbox{for}\quad
0\leq j<\infty.\label{semi}
\end{eqnarray}
To prove the proposition it is enough to check that the conditions
$(P_{-u})_{-v}=(P_{-v})_{-u}$, $(Q_{u})_{v}=(P_{v})_{u}$ hold only
together with the constraints (\ref{semibc}), (\ref{semi}). It is
clear that the system (\ref{Peqs}), (\ref{Qeqs}) is the matrix
form of the following system of the scalar equations
\begin{equation}\nonumber
\psi_u^j=e^{f_u^j-f^j}\psi^j-\psi^{j+1},\quad
y_{-v}^j=e^{f_{-u}^j-f^j}y^j-y^{j+1}, \quad\mbox{for}\quad j\geq0,
\end{equation}
\begin{equation}\nonumber
\psi_v^j=\psi^j +e^{f_v^j-f^{j-1}}\psi^{j-1},\quad
y_{-u}^j=y^j+e^{f_{-u}^j-f^{j-1}}y^{j-1}, \quad\mbox{for}\quad
j\geq1,
\end{equation}
\begin{equation}\nonumber
y^0_{-u}=y^0+X_{-u}\psi^0_{-u}, \quad \psi^0_{v}=\psi^0+H_vy^0_v,
\end{equation}
\begin{equation}\nonumber
Y_{-u}=z_{-u}Y+X_{-u,v}z_{-u}\Psi, \quad
\Psi_{v}=z_{-u}\Psi+H_{-u,v}z_{-u}Y,
\end{equation}
\begin{equation}\nonumber
Y_{-v}=y^0, \quad \Psi_{u}=\psi^0.
\end{equation}
In other words one has to check validity of the conditions
$(Y_{-u})_{-v}=(Y_{-v})_{-u}$, $(\Psi_{u})_{v}=(\Psi_{v})_{u}$,
$(y^j_{-u})_{-v}=(y^j_{-v})_{-u}$,
$(\psi^j_{u})_{v}=(\psi^j_{v})_{u}$. Moreover, it is enough to
take $j=0$, because for the other $j$ it is really true. The
conditions required are equivalent the following five equations:
\begin{eqnarray}
&&X_v=X\frac{e^{f_v-f}}{z},\nonumber\\
&&z^1e^{-\Delta f}=1+(z-1)\frac{z^1_{-u}}{z^{-1}_{-v}},\nonumber\\
&&X_v=X\frac{e^{f_v-f}}{z},\nonumber\\
&&z^1=ze^{\Delta f}, \nonumber\\
&&HX=\frac{z_{-v}-1}{z_{-v}}.\nonumber
\end{eqnarray}
The second and fourth of them are the consequences of the boundary
condition $z^1_v=z^{-1}_u$ and the others have already been proved
to be consistent under the boundary condition.

In order to find the continuum limit as $h\rightarrow0$ rewrite
the boundary condition $z^1_v=z^{-1}_u$ in a more explicit form
\begin{equation}\label{ebc}
f^{-2}_{u,-v}=f^{-1}+f-f^1_{-u,v}.
\end{equation}
Hence $f_u=f+hD_xf+O(h^2)$ and $f_v=f+hD_yf+O(h^2)$ then setting
$h\rightarrow0$ one gets $f^{-2}=f^{-1}+f-f^1$. Then it follows
from (\ref{toda}) that $D_xD_y(f+f^{-1})=0$. Consequently,
$f=-f_{-1}+a(x)+b(y).$ Remove functions $a(x)$ and $b(y)$ by the
dilatation $f=\tilde f+a(x)/2+b(y)/2$. As a result one gets the
boundary condition searched $f^{-1}=-f$.

Notice that when the additional constraint $f(u,v,k)=f(u+v,k)$ is
imposed on the chain (\ref{dtoda}), which reduces it to 1+1
dimensional discrete Toda chain, the boundary condition
(\ref{ebc}) is reduced to the form $f(u,-1)=-f(u,0)$ found earlier
by Yu.Suris in \cite{sur} (see also \cite{kaz}).

Consider now the other half of the chain which is located on the
left half-line $k\leq k_0$. Formulae (\ref{cons2}) allow one to
exclude from the Lax pair the functions $y^0$ and $\psi^0$, and
then rewrite the shifted variables $y^{-1}_{-v}$ and
$\psi^{-1}_{u}$ in the form
\begin{equation}\label{cons4}
y^{-1}_{-v}=y^{-1}e^{f^{-1}_{-v}-f^{-1}}-{1\over R}\psi^{-1}_{-v},
\quad \psi^{-1}_{u}=\psi^{-1}e^{f^{-1}_{u}-f^{-1}}-{1\over
S}y^{-1}_{u}.
\end{equation}
Introduce additional dynamical variables
$\Psi^{-1}:=\psi^{-1}_{-v}$ and $Y^{-1}:=y^{-1}_{u}$. Then the
previous equation can be transformed as follows
\begin{equation}\label{cons5}
y^{-1}_{-v}=y^{-1}e^{f^{-1}_{-v}-f^{-1}}-{1\over R}\Psi^{-1},
\quad \psi^{-1}_{u}=\psi^{-1}e^{f^{-1}_{u}-f^{-1}}-{1\over
S}Y^{-1}.
\end{equation}
Below we will use the following analog of Lemma 1.

{\bf Lemma 3.} Solutions of the equations (\ref{R}) and (\ref{S})
can be chosen to satisfy the constraint
\begin{equation}\label{RS}
R_{uv}S=\frac{z}{z-1}.
\end{equation}

To prove the lemma one has to express $H$ and $X$ through $R$ and
$S$ and substitute them into (\ref{HX}).

{\bf Lemma 4.}. The shifts $Y^{-1}_{-u}$, $Y^{-1}_{-v}$,
$Y^{-1}_{v}$, $\Psi^{-1}_{u}$, $\Psi^{-1}_{v}$, $\Psi^{-1}_{-u}$
of the variables $Y^{-1}$ and $\Psi^{-1}$ are linearly expressed
through the finite number of elements of the enlarged dynamical
set\\
$\{\Psi^{-1}, Y^{-1},\psi^{-1},y^{-1},\psi^{-2},y^{-2},...\}$:
\begin{eqnarray}
&& Y^{-1}_{-u}=y^{-1},\nonumber\\
&& Y^{-1}_{v}={1\over z}e^{f^{-1}_{uv}-f^{-1}_{u}}Y^{-1}+
{1\over R_{uv}} e^{f^{-1}_{uv}-f^{-1}}\psi^{-1},\nonumber\\
&& Y^{-1}_{-v}=z_{-v}e^{f^{-1}_{u,-v}-f^{-1}_{u}}Y^{-1}-
{z_{-v}\over R_{u}} e^{f^{-1}_{u,-v}-f^{-1}_{-v}}\Psi^{-1},\nonumber\\
&&\Psi^{-1}_{v}=\psi^{-1},\nonumber \\
&&\Psi^{-1}_{-u}={1\over z_{-u,-v}} e^{f^{-1}_{-u,-v}-f^{-1}_{-v}}
\Psi^{-1}+{1\over S_{-u,-v}}e^{f^{-1}_{-u,-v}-f^{-1}}y^{-1}, \nonumber\\
&& \Psi^{-1}_{u}=z_{-v}e^{f^{-1}_{u,-v}-f^{-1}_{-v}}
\Psi^{-1}-{z_{-v}\over
S_{-v}}e^{f^{-1}_{u,-v}-f^{-1}_{u}}Y^{-1}.\nonumber
\end{eqnarray}

Lemma 4 is proved similarly to the lemma 2.

Introduce infinite dimensional vectors-columns $\tilde{P}$ and
$\tilde{Q}$ as
\begin{equation}\label{tPtQ}
\tilde{P}=\left(%
\begin{array}{c}
 \cdots \\
 y^{-3}  \\
  y^{-2} \\
  y^{-1} \\
 Y^{-1}  \\
 \end{array}%
\right), \qquad
\tilde{Q}=\left(%
\begin{array}{c}
  \cdots \\
  \psi^{-3} \\
  \psi^{-2} \\
  \psi^{-1} \\
 \Psi^{-1} \\
\end{array}%
\right),
\end{equation}
and infinite dimensional matrices
\begin{equation}\label{tA}
\tilde{A}=\left(%
\begin{array}{ccccc}
\cdots & \cdots & \cdots & \cdots & \cdots \\
\cdots & 1 & 0 & 0 & 0 \\
\cdots & e^{f^{-2}_{-u}-f^{-3}}& 1& 0 & 0 \\
\cdots & 0 & e^{f^{-1}_{-u}-f^{-2}} & 1 & 0 \\
\cdots & 0 & 0 & 1 & 0 \\
\end{array}%
\right),
\end{equation}
\begin{equation}\label{tB}
\quad
\tilde{B}=\left(%
\begin{array}{ccccc}
  \cdots & \cdots & \cdots & \cdots & \cdots \\
\cdots & e^{f^{-3}_{-v}-f^{-3}} & -1 & 0 & 0 \\
\cdots &  0 & e^{f^{-2}_{-v}-f^{-2}} & -1 & 0 \\
\cdots &  0 & 0 & e^{f^{-1}_{-v}-f^{-1}} & 0\\
\cdots &  0 & 0 & 0 & z_{-v}e^{f^{-1}_{u,-v}-f^{-1}_{u}}\\
\end{array}%
\right)
\end{equation}
\begin{equation}\label{tC}
\tilde{C}=\left(%
\begin{array}{ccccc}
\cdots & \cdots & \cdots & \cdots & \cdots \\
\cdots &  1 & 0 & 0 & 0 \\
\cdots &  e^{f^{-2}_{v}-f^{-3}} & 1 & 0 & 0 \\
\cdots &  0 & e^{f^{-1}_{v}-f^{-2}} & 1 & 0 \\
\cdots &  0 & 0 & 1 & 0 \\
\end{array}%
\right), \quad
\end{equation}
\begin{equation}\label{tD}
\tilde{D}=\left(%
\begin{array}{ccccc}
\cdots & \cdots & \cdots & \cdots & \cdots \\
\cdots &  e^{f^{-3}_{u}-f^{-3}} & -1 & 0 & 0\\
\cdots &  0 & e^{f^{-2}_{u}-f^{-2}} & -1 & 0\\
\cdots &  0 & 0 & e^{f^{-1}_{u}-f^{-1}} & 0\\
\cdots &  0 & 0 & 0 & z_{-v}e^{f^{-1}_{u,-v}-f^{-1}_{-v}}\\
\end{array}%
\right)
\end{equation}
and two more matrices each with only two nonzero entries
\begin{equation}\label{tbd}
\tilde{b}=\left(%
\begin{array}{ccc}
\cdots & \cdots & \cdots \\
\cdots & 0 & 0 \\
\cdots & 0 & -{1\over R} \\
\cdots & 0 & -{z_{-v}\over R_u}e^{f^{-1}_{u,-v}-f^{-1}_{-v}}\\
\end{array}%
\right),\quad
\tilde{d}=\left(%
\begin{array}{ccc}
\cdots & \cdots & \cdots \\
\cdots & 0 & 0 \\
\cdots & 0 & -{1\over S} \\
\cdots & 0 & -{z_{-v}\over S_{-v}}e^{f^{-1}_{u,-v}-f^{-1}_{u}}\\
\end{array}%
\right).
\end{equation}

{\bf Proposition 4.} The following system of equations
\begin{eqnarray}
&\tilde{P}_{-u}=\tilde{A}\tilde{P},\quad &\tilde{P}_{-v}
=\tilde{B}\tilde{P}+\tilde{b}\tilde{Q},\label{tPeqs}\\
&\tilde{Q}_v=\tilde{C}\tilde{Q},\quad&\tilde{Q}_u=\tilde{D}
\tilde{Q}+\tilde{d}\tilde{P}\label{tQeqs}
\end{eqnarray}
is consistent if and only if the function $f=f(u,v,k)$ solves the
semi-infinite lattice
\begin{eqnarray}
&&e^{\Delta f}=\frac{z^{-1}_{u,-v}}{z},\label{tsemibc}\\
&&e^{\Delta f^{j}}=\frac{z^{j+1}}{z^j}, \quad \mbox{for}\quad
-\infty<j\leq -1.\label{tsemi}
\end{eqnarray}

\section{Finite chains}

Close the semi-infinite chain (\ref{semibc}), (\ref{semi}) by
imposing an additional (degenerate) boundary condition
$e^{f^{N+1}}=0$. The obtained chain coincides with (\ref{ctoda2}).
One can close also the Lax pair by setting $y^{N+1}=0$,
$\psi^{N+1}=0$. The Lax pair found can be represented in the
matrix form
\begin{eqnarray}
&P_{-u}=AP+aQ,\quad &P_{-v}=BP,\label{cpeqs}\\
&Q_v=CQ+cP,\quad&Q_u=DQ.\label{cqeqs}
\end{eqnarray}
with the following matrix coefficients of dimension $(N+2)\times
(N+2)$ :
\begin{equation}\label{A}
A=\left(%
\begin{array}{ccccccc}
  z_u & 0 & 0 & \cdots & 0 & 0 & 0\\
  0 & 1 & 0 &  \cdots &  0  & 0 & 0 \\
  0 & e^{f^1_{-u}-f} & 1 & \cdots & 0 & 0 & 0 \\
   \cdots & \cdots & \cdots & \cdots & \cdots & \cdots & \cdots\\
  0 & 0 & 0 &\cdots & e^{f^{N-1}_{-u}-f^{N-2}} & 1 & 0 \\
  0 & 0 & 0 & \cdots & 0 & e^{f^N_{-u}-f^{N-1}} & 1\\
\end{array}%
\right), \quad
\end{equation}
\begin{equation}\label{B}
B=\left(%
\begin{array}{cccccc}
  0 & 1 & 0 & \cdots & 0 & 0  \\
  0 & e^{f_{-v}-f} & -1 & \cdots & 0 & 0 \\
  0 & 0 & e^{f^1_{-v}-f^1} & \cdots & 0 & 0  \\
  \cdots & \cdots & \cdots & \cdots & \cdots & \cdots \\
  0 & 0 & 0 & \cdots & e^{f^{N-1}_{-v}-f^{N-1}} & -1 \\
    0 & 0 & 0 & \cdots & 0& e^{f^N_{-v}-f^N}  \\
\end{array}%
\right),
\end{equation}
\begin{equation}\label{C}
C=\left(%
\begin{array}{ccccccc}
  z_v & 0 & 0 & \cdots & 0 & 0 & 0\\
  0 & 1 & 0 & \cdots & 0 & 0 & 0\\
  0 & e^{f^1_{v}-f} & 1 & \cdots & 0 & 0 & 0\\
  \cdots & \cdots & \cdots & \cdots & \cdots & \cdots & \cdots \\
  0 & 0 & 0 & \cdots  & e^{f^{N-1}_{v}-f^{N-2}} & 1 & 0 \\
  0 & 0 & 0 & \cdots & 0 & e^{f^N_{v}-f^{N-1}} & 1  \\
\end{array}%
\right),
\end{equation}
\begin{equation}\label{D}
D=\left(%
\begin{array}{cccccc}
  0 & 1 & 0 & \cdots & 0 & 0 \\
  0 & e^{f_{u}-f} & -1 & \cdots & 0 & 0 \\
  0 & 0 & e^{f^1_{u}-f^1} & \cdots & 0 & 0 \\
  \cdots & \cdots & \cdots & \cdots & \cdots & \cdots \\
  0 & 0 & 0 & \cdots & e^{f^{N-1}_{u}-f^{N-1}} & -1  \\
    0 & 0 & 0 & \cdots & 0 & e^{f^N_{u}-f^N} \\
\end{array}%
\right).
\end{equation}
The matrices $a$ and $b$ are also of the dimension $(N+2)\times
(N+2)$. Their entries are all zero except the first two entries of
the first columns: $a_{00}=X_{-u,v}z_{-u}$, $a_{10}=X_{-u}$,
$c_{00}=H_{-u,v}z_{v}$, $c_{10}=H_{v}$.

If one imposes the condition (\ref{cons}) on both ends of the
chain, one gets the chain
\begin{eqnarray}
&&e^{\Delta f^{-1}}=\frac{z}{z_{-u,v}^1},\nonumber \\
&&e^{\Delta f^{j}}=\frac{z^{j+1}}{z^j}, \quad \mbox{for}\quad
0\leq j\leq N-1,\label{fintcd}\\
&&e^{\Delta f^{N}}=\frac{z^{N-1}_{u,-v}}{z^N}\nonumber
\end{eqnarray}
(see chain (\ref{tctoda2}) in Introduction). Its Lax pair is given
by
\begin{eqnarray}
&\hat{P}_{-u}=\hat{A}\hat{P}+\lambda\hat{a}\hat{Q},\quad
&\hat{P}_{-v}
=\hat{B}\hat{P}+\hat{b}\hat{Q},\label{fPeqs}\\
&\hat{Q}_v=\hat{C}\hat{Q}+\lambda^{-1}\hat{c}\hat{P},
\quad&\hat{Q}_u=\hat{D} \hat{Q}+\hat{d}\hat{P},\label{fQeqs}
\end{eqnarray}
where $\lambda$ is the spectral parameter and
\begin{equation}\label{fA}
\hat{A}=\left(%
\begin{array}{cccccccc}
  z_u & 0 & 0 & \cdots & 0 & 0&0&0\\
  0 & 1 & 0 &  \cdots &  0  & 0 &0&0\\
  0 & e^{f^1_{-u}-f} & 1 & \cdots & 0 & 0&0&0 \\
\cdots & \cdots & \cdots &
\cdots & \cdots & \cdots & \cdots & \cdots \\
0  & 0 &0&\cdots & 1 & 0 & 0 & 0 \\
0  & 0 &0&\cdots & e^{f^{N-2}_{-u}-f^{N-3}}& 1& 0 & 0 \\
0  & 0 &0&\cdots & 0 & e^{f^{N-1}_{-u}-f^{N-2}} & 1 & 0 \\
0  & 0 &0&\cdots & 0 & 0 & 1 & 0 \\
\end{array}%
\right), \quad
\end{equation}
\begin{equation}\label{fB}
\hat{B}=\left(%
\begin{array}{cccccccc}
  0 & 1 & 0 & \cdots & 0 & 0 & 0 \\
  0 & e^{f_{-v}-f} & -1 & \cdots  & 0& 0 & 0 \\
  0 & 0 & e^{f^1_{-v}-f^1} & \cdots  & 0 & 0 & 0 \\
\cdots & \cdots &
\cdots & \cdots & \cdots & \cdots & \cdots \\
  0 & 0 & 0 &\cdots & -1 & 0 & 0 \\
  0 & 0 & 0 &\cdots & e^{f^{N-2}_{-v}-f^{N-2}} & -1 & 0 \\
  0 & 0 & 0 &\cdots & 0 & e^{f^{N-1}_{-v}-f^{N-1}} & 0\\
  0 & 0 & 0 &\cdots & 0 & 0 & z^N_{-v}e^{f^{N-1}_{u,-v}-f^{N-1}_{u}}\\
\end{array}%
\right),
\end{equation}
\begin{equation}\label{fC}
\hat{C}=\left(%
\begin{array}{cccccccc}
  z_v & 0 & 0 & \cdots & 0 & 0 & 0 & 0\\
  0 & 1 & 0 & \cdots & 0 & 0 & 0 & 0\\
  0 & e^{f^1_{v}-f} & 1 & \cdots & 0 & 0 & 0 & 0\\
  \cdots & \cdots & \cdots & \cdots & \cdots & \cdots & \cdots & \cdots\\
  0 & 0 & 0 & \cdots & 1 & 0 & 0 & 0\\
  0 & 0 & 0 & \cdots & e^{f^{N-2}_{v}-f^{N-3}} & 1 & 0 & 0\\
  0 & 0 & 0 & \cdots & 0 & e^{f^{N-1}_{v}-f^{N-2}} & 1 & 0 \\
  0 & 0 & 0 & \cdots & 0 & 0 & 1 & 0  \\
\end{array}%
\right),
\end{equation}
\begin{equation}\label{fD}
\hat{D}=\left(%
\begin{array}{cccccccc}
  0 & 1 & 0 & \cdots & 0 & 0 & 0\\
  0 & e^{f_{u}-f} & -1 & \cdots  & 0 & 0 & 0 \\
  0 & 0 & e^{f^1_{u}-f^1} & \cdots  & 0 & 0 & 0\\
  \cdots & \cdots & \cdots & \cdots &  \cdots & \cdots & \cdots\\
  0 & 0 & 0 & \cdots  & -1 & 0 & 0 \\
  0 & 0 & 0 & \cdots  & e^{f^{N-2}_{u}-f^{N-2}} & -1  & 0 \\
  0 & 0 & 0 & \cdots & 0 & e^{f^{N-1}_{u}-f^{N-1}} & 0  \\
  0 & 0 & 0 & \cdots & 0 & 0 &
                           z^N_{-v}e^{f^{N-1}_{u,-v}-f^{N-1}_{-v}}\\
\end{array}%
\right),
\end{equation}
\begin{equation}\label{hac}
\hat{a}=\left(%
\begin{array}{ccccc}
  X_{-u,v}z_{-u} & 0 & \cdots & 0\\
  X_{-u} & 0 & \cdots  & 0\\
  0 & 0 & \cdots  & 0\\
  \cdots & \cdots & \cdots & \cdots\\
    0 & 0 & \cdots  & 0\\
\end{array}%
\right),\quad
\hat{c}=\left(%
\begin{array}{ccccc}
  H_{-u,v}z_{v} & 0 & \cdots & 0 \\
  H_{v} & 0 & \cdots & 0 \\
  0 & 0 & \cdots & 0 \\
  \cdots & \cdots & \cdots & \cdots\\
   0 & 0 & \cdots & 0 \\
\end{array}%
\right),
\end{equation}
\begin{equation}\label{hbd}
\hat{b}=\left(%
\begin{array}{ccccc}
0 & \cdots & 0 & 0 \\
\cdots & \cdots & \cdots & \cdots \\
0 & \cdots & 0 & 0 \\
0 & \cdots & 0 & -{1\over R^N} \\
0 & \cdots & 0 & -{z^N_{-v}\over R^N_u}e^{f^{N-1}_{u,-v}-f^{N-1}_{-v}}\\
\end{array}%
\right),\quad
\hat{d}=\left(%
\begin{array}{ccccc}
0 & \cdots & 0 & 0 \\
\cdots & \cdots & \cdots & \cdots \\
0 & \cdots & 0 & 0 \\
0 & \cdots & 0 & -{1\over S^N} \\
0 & \cdots & 0 & -{z^N_{-v}\over S^N_{-v}}e^{f^{N-1}_{u,-v}-f^{N-1}_{u}}\\
\end{array}%
\right),
\end{equation}
where $R^N$ and $S^N$ are defined through equations
\begin{eqnarray}
\label{RN} &R^N_u=R^Nz^N_{-v}e^{f^{N-1}_{u,-v}-f^{N-1}_{-v}},
\quad
&R^N_v=R^Nz^N_{-u}e^{f^N_{-u}-f^N_{-u,v}},\\
&S^N_u=S^N\frac{e^{f^N-f^N_u}}{z^{N-1}_{-u,-v}},\qquad\qquad
&S^N_v=S^N{\displaystyle{\frac{e^{f^{N-1}_v-f^{N-1}}}{z^{N-1}}}},
\label{SN}\\
&\label{RSN} R^N_{uv}S^N={\displaystyle{\frac{z^N}{z^N-1}}}.&
\end{eqnarray}

Explain how the spectral parameter has been introduced. Evidently
each of the functions $H$ and $X$ is defined by the equations
(\ref{H}), (\ref{X}) up to the constant multiplier. Due to the
constraint (\ref{HX}) imposed above, the only constant remains
free, and it is taken as the spectral parameter.

\paragraph{\large Acknowledgments.}
The work has been supported by grants RFBR\#04-01-00190 and
RFBR\#05-01-97910-r-agidel-a.

\end{document}